\documentclass{article}
\usepackage{arxiv}

\usepackage[utf8]{inputenc}
\usepackage[T1]{fontenc} 
\usepackage{hyperref}
\usepackage{url}
\usepackage{booktabs}
\usepackage{amsfonts}
\usepackage{nicefrac} 
\usepackage{microtype} 
\usepackage{lipsum}
\usepackage{float}
\usepackage{graphicx}

\title{A Hybrid Quantum enabled RBM Advantage: Convolutional Autoencoders For Quantum Image Compression and
Generative Learning}

\author{
  Jennifer Sleeman, John Dorband, Milton Halem \\
  Department of Computer Science and Electrical Engineering\\
  University of Maryland, Baltimore County\\
  Baltimore, MD 21250 USA\\
  \texttt{\{jsleem1,dorband,halem\}@umbc.edu} \\
  }

\begin{document}
\maketitle             
\begin{abstract}
Understanding how the D-Wave quantum computer could be used for machine learning problems is of growing interest. Our work, in particular, evaluates the feasibility of using the D-Wave as a sampler for machine learning.  We describe a hybrid system that combines a classical deep neural network autoencoder with a quantum annealing Restricted Boltzmann Machine (RBM) using the D-Wave. We evaluate our hybrid autoencoder algorithm using two datasets, the MNIST dataset and the MNIST Fashion dataset.  We evaluate the quality of this method by using a downstream classification method where the training is based on quantum RBM-generated samples. Our method overcomes two key limitations in the current 2000-qubit D-Wave processor, namely the limited number of qubits available to accommodate typical problem sizes for fully connected quantum objective functions and samples that are binary pixel representations. As a consequence of these limitations we are able to show how we achieved nearly a 22-fold compression factor of grayscale 28 x 28 sized images to binary 6 x 6 sized images with a lossy recovery of the original 28 x 28 grayscale images. We further show how generating samples from the D-Wave after training the RBM, resulted in 28 x 28 images that were variations of the original input data distribution, as opposed to recreating the training samples. We formulated an MNIST classification problem using a deep convolutional neural network that used samples from a quantum RBM to train the MNIST classifier and compared the results with an MNIST classifier trained with the original MNIST training data set, as well as an MNIST classifier trained using classical RBM samples. Our hybrid autoencoder approach indicates advantage for RBM results relative to the use of a current RBM classical computer implementation for image-based machine learning and even more promising results for the next generation D-Wave quantum system. \textbf{}Our method for compression and image mappings is not constrained to RBMs, the autoencoder part of this method could be coupled with other quantum-based algorithms.
\end{abstract}


\section{Introduction} 
The D-Wave 2000Q quantum annealing system is an adiabatic quantum system \cite{farhi2000quantum}, which is composed of a 2048-qubit processor.  The quantum annealing algorithm can be used to achieve a performance improvement for a number of optimization problems \cite{hashizume2015singular,neukart2017traffic,ushijima2017graph}. Recent attention has explored both the theoretical prospects of machine learning on the D-wave \cite{romero2017quantum,amin2018quantum,khoshaman2018quantum} and using the D-Wave for machine learning applications \cite{biamonte2017quantum,o2018nonnegative}.  Building on previous work, we describe a hybrid system that uses a deep convolutional autoencoder neural network that provides a way to compress images to be used for quantum machine learning.  We demonstrate this compression technique using a Restricted Boltzmann Machine (RBM) and the D-Wave 2000Q.  The D-Wave 2000Q is limited by the number of qubits available, which limits the overall size of the problem that can be embedded on the D-Wave. In addition, using the D-Wave for image-based sampling provides another complication in that only binary information can be sampled.  We overcome both of these issues in our work by using a classical deep convolutional autoencoder to provide a translation from images represented on the classical machine and image representation on the D-Wave.  In doing so, we are able to achieve compression currently from an image size of 28 x 28 grayscale to an image size of 6 x 6 binary and recover the original 28 x 28 grayscale (with some loss).  We show how we have been able to train a RBM using the D-Wave as the sampler and how sampling from the D-Wave provides enough low-level noise to extract new image variants of images.  We compare the results of a hybrid quantum RBM with a hybrid classical RBM using a downstream classification problem.  This approach for quantum compression and image mapping is not constrained to quantum RBMs, but could be coupled with other quantum-based algorithms.
 
\section{Background} 
RBMs have a long history dating back to the original paper introducing Boltzmann machines \cite{ackley1985learning}. Boltzmann machines are known for their intractability on classical machines \cite{salakhutdinov2009deep} due to the connectivity among units.  In addition to every unit in the visible layer being connected to every unit in the hidden layer, the Boltzmann machine also contains connections between units within the same layer.  RBMs are restricted in that there is no connectivity among units within a given layer.  An early example of an autoencoder, the RBM is a simple neural network.  It is distinct from other neural networks in that it is probabilistic and represents an undirected graphical model. 

Due to this property, RBMs can be used to learn the stochastic representation of its input, modeling the input distribution \cite{salakhutdinov2009deep}.  During training, the model parameters are changed so that the probability distribution fits the input data \cite{fischer2014training}.    

\begin{figure}[H]
\centering
\includegraphics[width=.3\columnwidth]{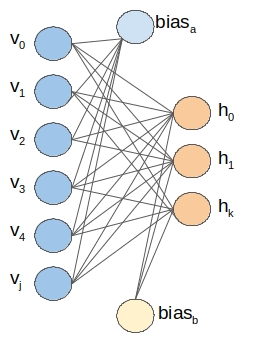}
\caption{Restricted Boltzmann Machine Network Diagram.}
\label{fig:rbm}
\end{figure}

The general structure of a RBM is one visible layer, one hidden layer, and two corresponding bias vectors as shown in Figure \ref{fig:rbm}. RBMs in the standard form have binary values for the $v$ visible and $h$ hidden units. 

\subsection{Energy Function} 
The energy of a given combination of $v$ and $h$ units is defined by the following equation:

\begin{equation}
E(v,h)=\sum_{i}a_{i}v_{i} - \sum_{j}b_{j}h_{j} - \sum \sum v_{i}w_{i,j}h_{j}
\end{equation}

where $v$ represents the visible units and $h$ represents the hidden units, $a$ and $b$ represent the bias, and $w$ represents the weight matrix and $E$ defines a probability distribution over $v$ and $h$ and is used for measuring the quality of the model by minimizing the $E$.

The joint probability distribution is defined in terms of the following equation (Gibbs distribution):
\begin{equation}
p(v,h)=\frac{e^{-E(v,h)}}{Z}
\end{equation}

$Z$ defines a partition function, which acts as a normalizer for this equation calculated over all possible states for $v$ and $h$.

\begin{equation}
Z=\sum_{v,h}\frac{e^{-E(v,h)}}{Z}
\end{equation}

Hidden and visible unit states are independent and calculated using conditional probabilities where the conditional probability of the visible units state is conditioned upon the current state of the hidden units, and conditional probability of the hidden units state is conditioned upon the current state of the visible units.  This independence is due to the that fact that there are no connections between units of a given layer. This improves the Gibbs sampling method because the states for a given layer can be jointly sampled \cite{fischer2014training}.

\subsection{Learning}
Maximizing the probability of the training data for the model can be compared to the maximum likelihood estimation, where a likelihood function is maximized by some configuration of the state space.  To maximize the likelihood function, the gradients of the log-likelihood are needed.  Calculating these gradients are not tractable.  Since gradient calculations are intractable, Gibb's sampling \cite{carter1994gibbs}, a Markov Chain Monte Carlo (MCMC) method, is used to sample from the joint Boltzmann distribution. Often the Gibb's sampling requires many steps and can become computationally expensive. Therefore, a method known as Contrastive Divergence \cite{hinton2002training} can be used to perform only $n$ steps of the Gibb's sampling.  It has been shown that even in some cases $1$ step is enough for training the RBM \cite{hinton2002training}.

Given a set of training samples the RBM is trained to learn how to adjust model state such that the probability distribution is fit to the training data probability distribution.

\subsection{Generative Models}
\begin{figure}[H]
\centering
\includegraphics[width=.90\columnwidth]{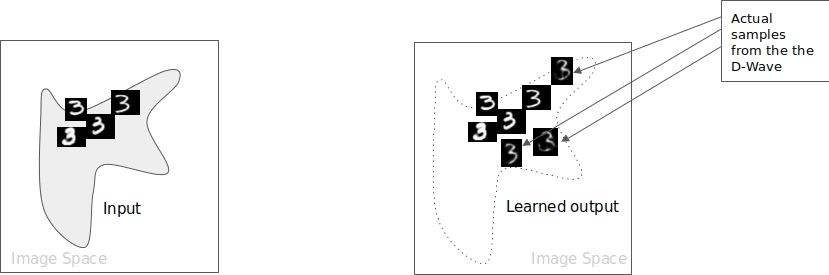}
\caption{Generative Learning Data Distributions.}
\label{fig:generative}
\end{figure}

Generative models model the joint probability distribution rather than a conditional probability distribution.  For image generation, deep networks are used to learn a distribution that is similar to the input data distribution.  The distribution of sampled output does not have a relationship with the distribution of samples from input variables.  RBMs learn a joint probability distribution $P(v,h)$, where $v$ represents the visible units and $h$ represents the hidden units.  Given $P(v,h)$, sampling from this distribution, could enable generated output that is not necessarily a recreation of a sample from the input distribution.  Previous work has used RBMs for generative sampling \cite{taylor2007modeling,schmah2009generative,sukhbaatar2011robust}. We illustrate this type of learning in Figure \ref{fig:generative}.

\subsection{RBMs on the D-Wave}
In order to implement a RBM which uses the D-Wave, the RBM problem to be solved on the D-Wave needs to be expressed as a QUBO objective function \cite{boros2007local}.  We express the QUBO in terms of a chimera graph \cite{cai2014practical} which is the architecture formed by D-Wave qubit connectivity.  This architecture (a 2-D grid) entails groupings of 'unit cells' that are connected, where each cell contains a set of qubits that have bipartite connectivity locally and is connected to qubits in other sets through couplers\cite{dwavedocs}.  Minor embedding entails mapping logical qubits to physical qubits where nodes and edges map to logical qubits and couplers in the chimera graph.  This architecture is similar to the architecture of a RBM.  

Bias noise in this model are different than bias noise in classical RBMs, in that quantum RBM bias variables are random \cite{dumoulin2014challenges}.

We follow the approach of mapping images to the RBM by having each pixel of the image represented by a visible unit of the RBM.  However, since images go through the autoencoder before RBM processing, we formulate an encoding based on the number of pixels and a compression size.

\section{Related Work} 
Early work by Dorband \cite{dorband2015boltzmann} explored a RBM implementation using the D-Wave.  This implementation used a different approach and was not necessarily identifying generative differences between a classical and quantum approach. Generative sampling using the D-Wave and a similar approach for the RBM, in the past yielded poor results. In the  work by Thulasidasan et al.   \cite{thulasidasan2016generative}, generative sampling after training MNIST down-scaled samples were not visibly distinguishable.  Previous work by Adachi et al. \cite{adachi2015application} described a generative RBM with 32 visible nodes and 32 hidden nodes using 512 qubits for sampling.  They showed that the quantum sampling was able to achieve comparable accuracy to the classical system with fewer iterations.

Work by Romero et al. \cite{romero2017quantum} describe the need for tools which reduce experimental overhead as advantageous. They describe the idea of a quantum autoencoder.  Though they describe using the quantum autoencoder for compression, they solve a different problem than what we propose.
Work by Ni et al. \cite{Ni2018} performed a comparison between a classical RBM and a quantum RBM using binary problems and saw some improvements using the quantum RBM.  Most experiments were simple binary problems.  We build on this work and extend it to support RBMs for real problems (such as MNIST).  
Recent work by Amin et al. \cite{amin2018quantum} explored using a quantum RBM for generative sampling.  Interestingly, when experimenting with their model of fully-connected 8 inputs and 3 outputs of binary data, they showed that the distribution learned using the quantum method when compared with the classical for a small test was very different from the actual distribution.  
Recent work by Khoshaman et al. \cite{khoshaman2018quantum} used a quantum Boltzmann machine to generate the latent space for a Variational autoencoder, showing state of the art results using the MNIST dataset. This work is most closely related to our work in that we both explore using the D-Wave for sampling to generate latent space.  However, our approach includes providing a hybrid classical quantum approach to overcome quantum hardware limitations that affect the number of qubits available to represent problems.
 
To summarize, many of the early theoretical contributions have shown that using the D-Wave for generative sampling can enable faster learning of the latent space. However, as Amin et al. \cite{amin2018quantum} concluded, the learned distribution can deviate significantly from a given actual distribution.  We believe the classical autoencoder performing the translation between the classical system and the quantum system acts as a stabilizer for the latent space since part of the latent space is captured on the classical side.

\section{Approach} 
\begin{figure}[H]
\centering
\includegraphics[width=1.0\columnwidth]{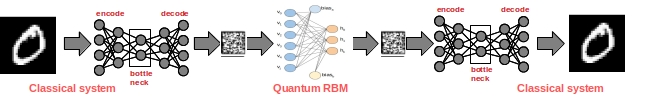}
\caption{Hybrid Approach that used a Classical Autoencoder to map the Image Space to a Compressed Space.}
\label{fig:approach}
\end{figure}

Our approach is designed to address the challenges of image sampling using the D-Wave with regards to the number of qubits available by simultaneously mapping binary output to floating point values and the original image space to a compressed binary image space.  The overall architecture of our approach is shown in Figure \ref{fig:approach} using the MNIST dataset as an example dataset.  An autoencoder is trained on the original input, a bottleneck is defined in terms of the compression size, and the autoencoder learns how to map binary compressed encodings to the original grayscale full resolution images. 

We use a 3-layer convolutional autoencoder with compressions sizes of 6 x 6  and 7 x 7.  We use the bottleneck of the autoencoder to generate the binary compression.

After training, we take the encoded data and use it to train the RBM.  Each pixel of the encoded input is represented by a unit in the visible layer.  For example, a 6 x 6 binary compression would have 36 units in the visible layer.  To create a mapping of this for use on the D-Wave we create an embedding which has 36 visible units, 18 hidden units, and connectivity from each visible unit to each hidden unit.  Hence, with 2048 qubits available (not all are available) on the D-Wave 2000Q system, if we compress larger than a 7 x 7 size we go beyond the capacity of the D-Wave.

During training, the RBM learns the best configuration for recreating the binary encodings, meaning the learned data distribution moves towards the actual training data distribution.  When this learned distribution no longer improves (converges), we use the trained RBM for sampling binary encodings.  Those encodings are then decoded by the autoencoder to obtain images in the original representation space, both in size and in pixel values (grayscale rather than binary).

\section{Experimental Setup} 
We used the MNIST dataset to evaluate this approach.  The autoencoder runs on a classical machine (Intel Core i7-7700HQ CPU 2.8GHz x 8) with 32 GBs of memory, using a GeForce GTX 1060 GPU and takes as input grayscale MNIST digits at a size of 28 x 28. The output from the encoder is compressed binary data of a size 7 x 7 in run 1 and 6 x 6 in run 2 as binary encodings.  The autoencoder is trained using the 60,000 MNIST training digits.  The RBM is then trained using the binary encodings of these 60,000 digits.  Using the D-Wave during training, the model is updated by sampling states of $P(v,h)$ that minimize $E$. Once convergence is reached during training, i.e. samples can be recreated, the training is ended.  The D-Wave is then used for sampling to obtain learned binary encodings.  The samples obtained from the D-Wave are then given to the classical autoencoder to recover 28 x 28 sized grayscale MNIST images.  To assess the quality of the MNIST generated images, we performed downstream classification experiments.

For the quantum RBM, we use the D-Wave API for working with the quantum annealer.  We use the D-Wave 2000Q solver.  We create a BinaryQuadraticModel for the QUBO and use minorminer to create the embedding. Samples are obtained using the DWaveSampler.  These experiments do not include the new D-Wave Hybrid API (our experimentation began before the Hybrid API was available).  The classical RBM is built to be as similar as possible to the quantum RBM.  Both methods use Gibbs sampling and contrastive divergence.  Both have a visible layer with unit size equal to the number of encoded values (36 for 6 x 6 compressions and 49 for 7 x 7 compressions) and a hidden layer that is half of the size of the visible layer.

\section{Experimental Results} 
After training the quantum RBM we sample from the D-Wave to obtain 60,000 samples, where each class is balanced based on the original size of the MNIST training data.  We use those samples to train a downstream deep convolutional neural network to classify 10,000 unseen MNIST test digits.  We compare these results with training a downstream deep convolutional neural network using the original MNIST training data and classify the same 10,000 unseen MNIST test digits.  We also compare these results with training a downstream deep convolutional neural network using the classical RBM recreated samples as training data and classify the same 10,000 unseen MNIST test digits.  With the classical RBM Gibbs sampling is used. 

Though classical RBMs and quantum RBMs are different in the way they learn, this comparison provides insights into how the D-Wave sampling method compares to the Gibb's sampling method performed on the classical system.  These experiments are intended to be the overfit case, in that we train on a set of images and try to regenerate those training samples. 

The training data in each experiment consists of 60,000 samples (with a subset reserved for validation) and the test set consists of 10,000 samples (unseen).  We trained the classifier for 5 epoch, since accuracy can reach $~100\%$ with the original data set in 5 epoch. 

As shown in Table \ref{tab:results_1}, the classifier that was trained with the original MNIST training data, achieved about a $99\%$ accuracy (ExpID 1).  This measure serves as a benchmark for maximum accuracy that could be achieved.  We then used samples from a classical RBM after training it on the original 28 x 28 grayscale MNIST digits.  The learned samples were then used to train the downstream MNIST classifier.  In this case, the results were $~98\%$ (ExpID 2) as shown in Table \ref{tab:results_1}.   We used to get a baseline for what the classical RBM trained on the original data could achieve.

\begin{table}[H]
\centering
\begin{tabular}{|p{1.5cm}|p{8.5cm}|p{2cm}|}
\hline
\textbf{ExpID} &
\centering\textbf{Training Data Variants} & \textbf{Accuracy On Test Set} \\\hline
 1 & Original MNIST 28 x 28 digits (baseline) &  0.99  \\\hline
 2 & Sampled from Classical RBM  MNIST 28 x 28 digits (no encodings)  & 0.98 \\\hline
\end{tabular}
\caption{Comparing MNIST classification accuracy scores when using the original training data and samples generated from a classical RBM.}
\label{tab:results_1}
\end{table}

The next set of experiments as shown in Table \ref{tab:results_2} was used to evaluate the classical autoencoder binary compression recovery of the original MNIST data representation.

\begin{table}[H]
\centering
\begin{tabular}{|p{1.5cm}|p{8.5cm}|p{2cm}|}
\hline

\textbf{ExpID} &
\centering\textbf{Training Data Variants} & \textbf{Accuracy On Test Set} \\\hline
 3 & Encoded/Decoded Binary Translated MNIST 28 x 28 digits (no RBM, no compression, just binary to grayscale mappings) &  0.97\\\hline
 4 & Encoded/Decoded Binary MNIST 16 x 16  decoded to 28 x 28 digits (no RBM) &  0.95\\\hline
 5 & Encoded/Decoded Binary MNIST 6 x 6 decoded to 28 x 28 digits (no RBM) &  0.91\\\hline
\end{tabular}
\caption{Comparing MNIST classification accuracy when using different compressed encodings that are decoded to recover the original MNIST representations.}
\label{tab:results_2}
\end{table}

ExpID 3 was used to evaluate MNIST digits that were converted from grayscale to binary then to grayscale using the original 28 x 28 size.  ExpID 3 was used to evaluate compression only, the method of converting from grayscale to binary and back to grayscale.  The results in this case were $~97\%$ accuracy (averaged over three runs).  This was compared to compressing the digits to a binary 16 x 16 size then decoded back to 28 x 28 grayscale (ExpID 4).  When using these images for training the classifier it achieved a $~95\%$ accuracy.  The importance of a 16 x 16 compression is that this provides an anticipated size that the next generation D-Wave will support given an increase in the number of qubits that will be available.  We treat this accuracy as a potential upper bound for what could be achieved when a 16 x 16 sized compression could be supported on the D-Wave.  ExpID 5 evaluated MINST digits compressed to a binary 6 x 6 size and decoded back to the 28 x 28 grayscale representation, achieving a $~91\%$ accuracy.  These encoding/decoding results provide a secondary baseline for results, in that they provide an upper bound for accuracy that could be obtained using encoded learned samples from a RBM.  

\begin{figure}[H]
\centering
\includegraphics[width=1.0\columnwidth]{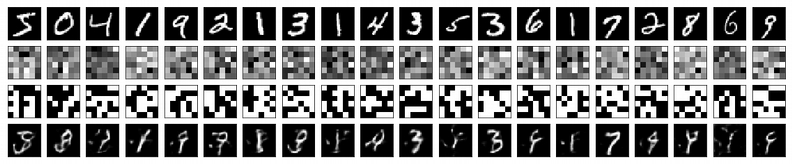}
\caption{Examples of 6 x 6 binary encoding and decoding to recover 28 x 28 grayscale MNIST digits.}
\label{fig:28x28encodingdecoding}
\end{figure}

Examples of 6 x 6 grayscale encoding/decoding results are shown in Figure \ref{fig:28x28encodingdecoding}, where the first row represents the original 28 x 28 MNIST digits, the second row represents the  6 x 6 grayscale encodings, the third row represents the 6 x 6 binary encodings and the fourth row represents the recovered 28 x 28 digits.
It is observed from the decoded digits that there are times when the digits are incorrectly decoded in the case of 6 x 6.  This is due to the loss incurred when compressing from the 28 x 28 to 6 x 6 binary.

In Table \ref{tab:results_3} we show results related to using MNIST digits recovered from the encodings of the size required for embedding the RBM model on the D-Wave.

\begin{table}[H]
\centering
\begin{tabular}{|p{1.5cm}|p{8.5cm}|p{2cm}|}
\hline

\textbf{ExpID} &
\centering\textbf{Training Data Variants} & \textbf{Accuracy On Test Set} \\\hline
 6 & Encoded/Decoded Binary MNIST 6 x 6 to translated to 28 x 28 digits classical RBM using Gibbs sampling & 0.75\\\hline
 7 & Encoded/Decoded Binary MNIST 6 x 6 to translated to 28 x 28 digits RBM using the D-Wave as a sampler  & 0.72\\\hline
\end{tabular}
\caption{Comparing MNIST classification accuracy when using classical and quantum RBM samples.}
\label{tab:results_3}
\end{table}

We compress MNIST digits from 28 x 28 to binary 7 x 7 and also 28 x 28 to binary 6 x 6. In these experiments, we encoded and converted the MNIST digits to a 6 x 6 binary representation.  We then trained a purely classical RBM and also the quantum RBM.  The number of visible layers was composed of $36$ units.  The number of hidden layers was composed of $18$ units for both RBMs.  We set the learning rate for the classical RBM to be the same as the quantum RBM.

We achieved a 75\% accuracy using the downstream classifier with the samples obtained Gibbs sampling using the classical RBM (ExpID 6).  We achieved a 72\% accuracy using the downstream classifier with the samples obtained from the quantum RBM (ExpID 7). By modifying the autoencoder to include dropout layers and by increasing the RBM epoch, we were able to achieve a 12\% increase in downstream classification results which gave us a 72\% accuracy (previously we achieved 60\% accuracy on average).

\begin{figure}[H]
\centering
\includegraphics[width=1.0\columnwidth]{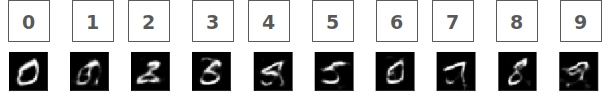}
\caption{Recovered MNIST digits from the quantum RBM after a 6 x 6 binary encoding.}
\label{fig:digitsfromdwave_6x6}
\end{figure}

For the quantum RBM, using 6 x 6 binary encodings, we were able to recover MNIST digits, as shown in Figure \ref{fig:digitsfromdwave_6x6} when sampling from the D-Wave after training and using the classical autoencoder to decode sampled encodings.

\begin{figure}[H]
\centering
\includegraphics[width=1.0\columnwidth]{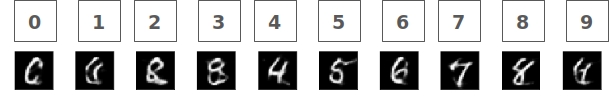}
\caption{Recovered MNIST digits from the quantum RBM after a 7 x7 binary encoding.}
\label{fig:digitsfromdwave_7x7}
\end{figure}

The quality of the digits tend to have a better appearance when using 7 x 7 binary encodings sampled from the D-Wave and decoded using the classical autoencoder, as shown in Figures \ref{fig:digitsfromdwave_7x7}. Though downstream classification results were not significantly different with modest improvements using 7 x 7 decoded samples. 

All results are shown in Table \ref{tab:all_results}.
\begin{table}[H]
\centering
\begin{tabular}{|p{1.5cm}|p{8.5cm}|p{2cm}|}
\hline
\textbf{ExpID} &
\centering\textbf{Training Data Variants} & \textbf{Accuracy On Test Set} \\\hline
 1 & Original MNIST 28 x 28 digits (baseline) &  0.99  \\\hline
 2 & Sampled from Classical RBM  MNIST 28 x 28 digits (no encodings)  & 0.98 \\\hline
 3 & Encoded/Decoded Binary Translated MNIST 28 x 28 digits (no RBM) &  0.97\\\hline
 4 & Encoded/Decoded Binary MNIST 16 x 16  to translated to 28 x 28 digits (no RBM) &  0.95\\\hline
 5 & Encoded/Decoded Binary MNIST 6 x 6 to translated to 28 x 28 digits (no RBM) &  0.91\\\hline
 6 & Encoded/Decoded Binary MNIST 6 x 6 to translated to 28 x 28 digits classical RBM using Gibbs sampling  & 0.75\\\hline
 7 & Encoded/Decoded Binary MNIST 6 x 6 to translated to 28 x 28 digits RBM using the D-Wave as a sampler  & 0.72\\\hline

\end{tabular}
\caption{Downstream classification results for all ExpIDs.}
\label{tab:all_results}
\end{table}

\subsection{Measuring Image Similarity}
To understand how samples generated using the quantum RBM differ from the original data set and from the classical RBM samples, we use a metric to measure image structural similarity \cite{wang2004image} defined by the following equation:
\begin{equation}
 SSIM(x,y) = \frac{(2\mu_{x}\mu_{y} + c_{1})(2\sigma{xy} + c_{2})}{(\mu^{2}_{x} + \mu^{2}_{y}+c_{1})(\sigma^{2}_{x} + \sigma^{2}_{y}+c_2)}
\end{equation}
Where $x$,$y$ are the two images to be compared and $\mu_{x}$ is the average of $x$, $\mu{y}$ is the average of $y$, $\sigma^{2}_{x}$ is the variance of $x$, $\sigma^{2}_{y}$ is the variance of $y$, and $\sigma{xy}$ is the correlation coefficient of $x$ and $y$.

We use this method to compare MNIST generated images to the original training data set of 60000 samples.  We also use this measure to measure how much similarity there is among samples within a dataset.  To use SSIM for the whole dataset would require comparing each image against 60,000 images.  This computation would be extremely computer intensive.  Instead, given we have a dataset,  $mnist_{generated}$, we randomly select $n$ set of images from $mnist_{generated}$ and compare each image with the rest of the $59,999$ images and average the results to calculate the SSIM.  In the case of measuring the SSIM of $mnist_{generated}$  and $mnist_{real}$, if we have 1-to-1 mapping between $mnist_{generated}$  and $mnist_{real}$, we compare them 1-to-1.  If we do not, have a 1-to-1 mapping, as in the case with the D-Wave generated samples, we use the same sampling method.

To evaluate what is the best $n$, we experimented with different values for $n$.  In Table \ref{tab:ssim_sampling}, we show the averege SSIM score given the number of samples used to calculate it for a given dataset of size 60,000 samples. Given 10 samples, 100 samples, or 1000 samples,  a stable average of how much similarity there is across digits is consistent.  Therefore, we use 100 samples to measure similarity among images in a given dataset.  
\begin{table}[H]
\centering
\begin{tabular}{|p{6cm}|p{4cm}|p{3cm}|}
\hline
\centering\textbf{Dataset} & \textbf{Average Image Sample Similarity score} & \textbf{Number of Samples to Compare} \\\hline
Original MNIST 28 x 28 digits &
0.5604 & 10 \\\hline
Encoded/Decoded Binary MNIST 7 x 7 to translated to 28 x 28 digits (no RBM)
& 0.5456 & 10 \\\hline
Original MNIST 28 x 28 digits &
0.5664 & 100  \\\hline
Encoded/Decoded Binary MNIST 7 x 7 to translated to 28 x 28 digits (no RBM)
& 0.5540 & 100 \\\hline
Original MNIST 28 x 28 digits &
0.5633 & 1000  \\\hline
Encoded/Decoded Binary MNIST 7 x 7 to translated to 28 x 28 digits (no RBM) &
0.5526 & 1000 \\\hline
\end{tabular}
\caption{Measuring how similar a sample of decodings is to the remaining decodings in the dataset after binary encoding/decoding and comparing this to measuring similarity among the original MNIST digits.}
\label{tab:ssim_sampling}
\end{table}

We compare this measure for the original MNIST dataset with digits that are decoded after encoding and compressing down to a 7 x 7 size.  For 7 x 7 after binary encoding and decoding to the original grayscale 28 x 28 representation, image sample similarity scores shows only a .01 difference from the original MNIST to the encoded/decoded recovered digits.  In Table \ref{tab:ssim_sampling_2} we show a comparison of image sample similarity scores for the original MNIST dataset, the encoded/decoded recovered digits, and the classical RBM learned encoded/decoded recovered digits.  As observed, there tends to be more duplication among what is learned using the classical RBM.

\begin{table}[H]
\centering
\begin{tabular}{|p{6cm}|p{6cm}|}
\hline
\centering\textbf{Dataset} & \textbf{Average Image Sample Similarity score (Using 100 samples)}  \\\hline
Original MNIST 28 x 28 digits &
0.5664\\\hline
Encoded/Decoded Binary MNIST 7 x 7 to translated to 28 x 28 digits (no RBM)
& 0.5540 \\\hline
 Encoded/Decoded Binary MNIST 7 x 7 to translated to 28 x 28 digits classical RBM using Gibbs sampling & 0.7362  \\\hline
  Encoded/Decoded Binary MNIST 7 x 7 to translated to 28 x 28 digits quantum RBM using D-Wave sampling & 0.589  \\\hline
\end{tabular}
\caption{Measuring dataset similarity using SSIM and comparing the original MNIST dataset with the binary encoded/decoded dataset, with the Classical RBM learned binary encoded/decoded dataset, and with the Quantum RBM learned binary encoded/decoded dataset.}
\label{tab:ssim_sampling_2}
\end{table}

To get a better ideas of how much images overlap structurally for a given class, we calculated SSIM measures for each class of the original training data set, again using 100 sized sample sets.  The digits which are classified as the number $1$ tend to have a higher average SSIM score, as would be expected.  We show these results in Table \ref{tab:ssim_sampling_3}.

\begin{table}[H]
\centering
\begin{tabular}{|p{2cm}|p{4cm}|p{6cm}|}
\hline
\centering\textbf{Digit} & \textbf{Total Size} & \textbf{Average Image Sample Similarity score}   \\\hline
0 & 5923 & 0.5786\\\hline
1 & 6742 & 0.7801\\\hline
2 & 5958 & 0.5812 \\\hline
3 & 6131 & 0.6032 \\\hline
4 & 5842 & 0.6139\\\hline
5 & 5421& 0.5949\\\hline
6 & 5918 & 0.6156 \\\hline
7 & 6265 & 0.6584 \\\hline
8 & 5851 & 0.6130 \\\hline
9 & 5949 & 0.6493 \\\hline
\end{tabular}
\caption{Measuring dataset similarity using SSIM by Digit using the original MNIST dataset.}
\label{tab:ssim_sampling_3}
\end{table}

\subsection{Generating MNIST Samples}
\begin{figure}[H]
\centering
\includegraphics[width=.60\columnwidth]{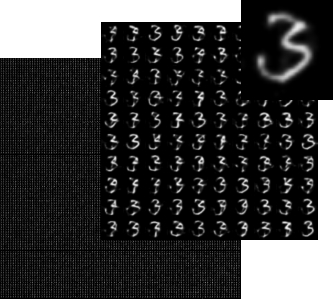}
\caption{D-Wave samples of the digit 3 after training the RBM.}
\label{fig:the_threes}
\end{figure}

We examined individual MNIST digits of large samples taken from the D-Wave.  Using the MNIST digit 3, we show a large sampling in Figure \ref{fig:the_threes} of this digit after training the RBM using the D-Wave as the sampler.  We sampled 100,000 3's. 

\begin{figure}[H]
\centering
\includegraphics[width=.7\columnwidth]{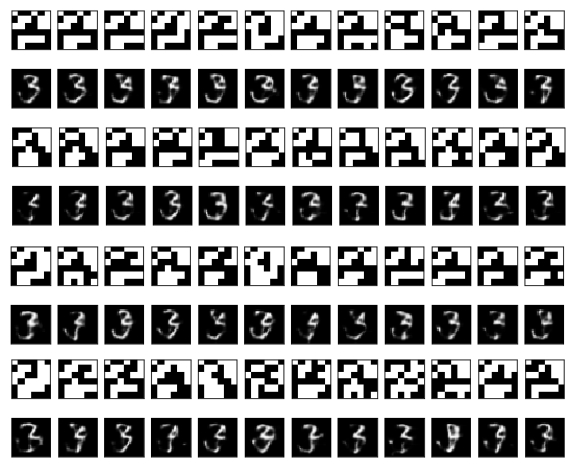}
\caption{D-Wave samples of the digit 3 after training the RBM.}
\label{fig:threes_binary}
\end{figure}

We also show its binary output and recovered digit in Figure \ref{fig:threes_binary} from the D-Wave sampling.  As can be seen, both binary output and recovered digits represent a distinct set of 3's that were not necessarily part of the original training distribution.

\begin{figure}[H]
\centering
\includegraphics[width=.4\columnwidth]{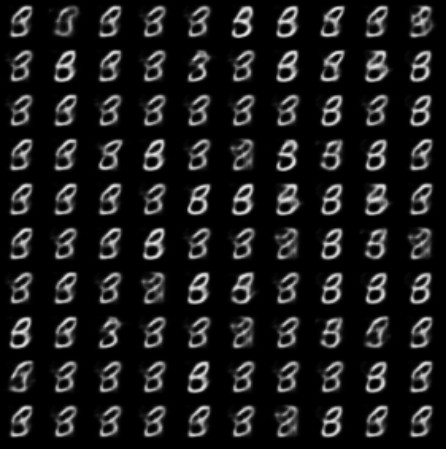}
\caption{Using Low Noise D-Wave Quantum Annealing for RBM sampling training on the digit 3.}
\label{fig:threes_low_noise}
\end{figure}

We were able to produce these variations in samples trained on 100's of samples of the original training digits.  Reproducing these variations in digits was repeatable on the original D-Wave 2000Q system.  When the D-Wave 2000Q was replaced with the D-Wave 2000Q lower-noise system, (holding all parameters constant), we were not able to reproduce these results, as shown in Figure \ref{fig:threes_low_noise}.  

\begin{figure}[H]
\centering
\includegraphics[width=.4\columnwidth]{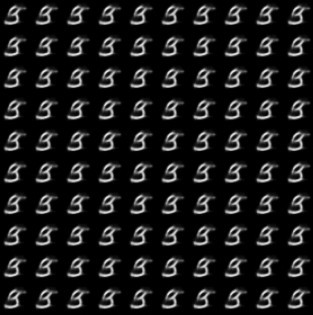}
\caption{Using Simulated Annealing for RBM sampling training on the digit 3.}
\label{fig:threes_sa}
\end{figure}

In addition, using simulated annealing for sampling also produced results that were not comparable shown in Figure \ref{fig:threes_sa}.  We conclude from these results that there was enough thermal temperature fluctuations that enabled the D-Wave sampling to result in variations in sampled encodings, and digits after decoding the generated encodings.  

However, with the D-Wave 2000Q lower-noise system, when we reduce learning rates and increased the number of epoch, we were able to achieve variations in samples. 

\subsection{The First Quantum Generated Fashion}
\begin{figure}[H]
\centering
\includegraphics[width=.7\columnwidth]{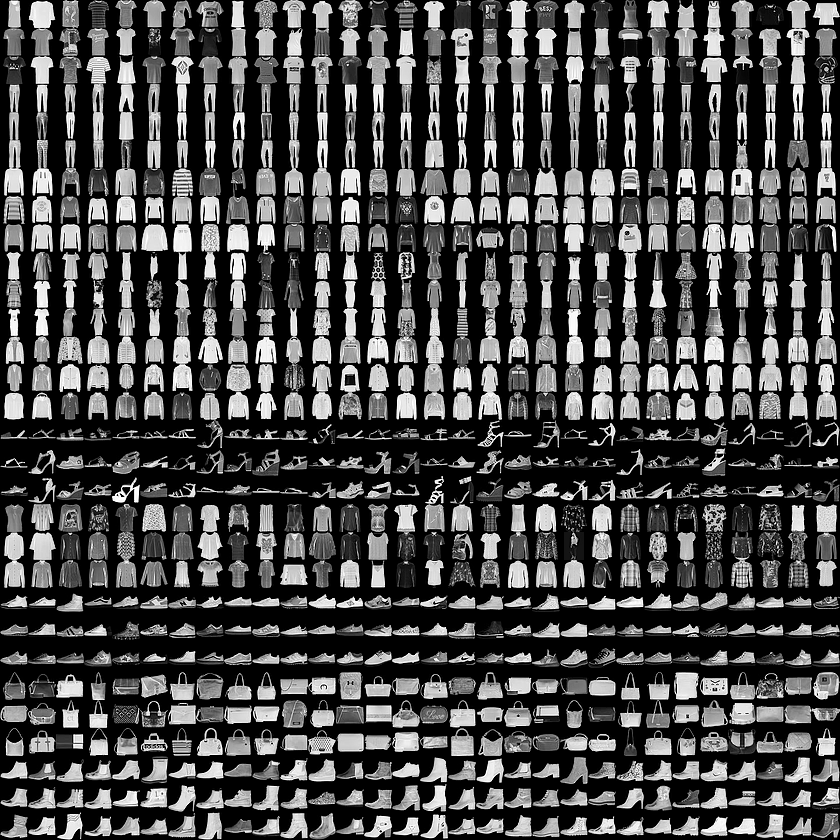}
\caption{Fashion MNIST}
\label{fig:fashion1}
\end{figure}

In addition to the MNIST dataset, we also trained the quantum RBM using the MNIST-Fashion dataset \cite{xiao2017fashion}.  Designed to be similar to the MNIST dataset as shown in Figure \ref{fig:fashion1}, but harder to classify, the Fashion MNIST dataset provides another dataset for experimenting with the D-Wave.  There are 10 classes, 60,000 images in total, and the images are grayscale sized at 28 x 28.

\begin{figure}[H]
\centering
\includegraphics[width=1\columnwidth]{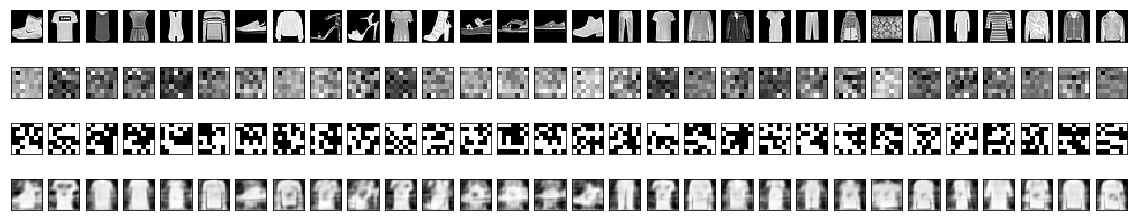}
\caption{Fashion MNIST Generated By Sampling the D-Wave After Training the Quantum RBM and Decoding the 7 x 7 Samples.}
\label{fig:fashion2}
\end{figure}

We show in Figure \ref{fig:fashion2} the first row is the original grayscale fashion images, the second row is the 7 x 7 encoded fashion, the third row is the 7 x 7 binary encoded fashion and the final row is the 28 x 28 grayscale decoded fashion.  As can be observed, the autoencoder tends to have a harder time generating details on shirts, shoes, and handbags.

\begin{figure}[H]
\centering
\includegraphics[width=.8\columnwidth]{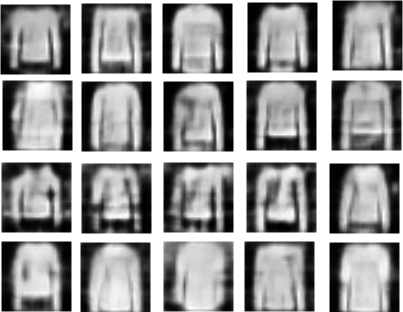}
\caption{New Shirts - Fashion MNIST Generated By Sampling the D-Wave After Training the Quantum RBM and Decoding the 7 x 7 Samples.}
\label{fig:fashion3}
\end{figure}

In Figure \ref{fig:fashion3} we show decoded samples from the D-Wave after training the quantum RBM.  Future experiments will include classification of generated Fashion MNIST samples.

\section{Observations and Conclusions}
The method we described provides a way to overcome the limitations of the D-Wave 2000Q by providing a hybrid method from mapping original data representations to a representation that could be processing using quantum annealing.  

Our prior attempts at using the classical autoencoder for this work were unsuccessful where learned encodings from the D-Wave did not decode to digits.  Initially we began with a fully connected autoencoder, which showed only hints of a digit recovered.  We saw improvement when moving to a convolutional neural network for the autoencoder.  However, the real improvements came when using a denoising autoencoder for this work.  By applying Gaussian noise to the images prior to encoding them, we saw improved quality when decoding the quantum sampled learned encodings.  

Using a classical RBM  with Gibb's sampling did not produce the same sort of variations as when the D-Wave was used as the sampler.  In fact, when running a number of experiments varying the epoch and learning rate on the classical RBM, we often saw the network learning only one digit and we saw this more frequently using 6 x 6 compressed encodings.  As we moved up to 16 x 16 encoding we could finally see the classical RBM learning different types of digits.  We conclude from these experiments that the quantum RBM was able to tolerate and learn from these highly compressed encodings whereas the classical RBM could not.  Though we are able to decode results from the classical RBM, they often decoded to same digit.  

Though RBMs are generative, using a RBM to generate new images from a latent representation, as opposed to recreating training samples, is not typically performed on the classical computer but has been achieved, for example with deep layered RBMs \cite{hu2016deep}. By taking advantage of the inherent noise on the D-Wave and the natural quantum properties of the D-Wave, we have been able to successfully use it to generate images. By generating images from sampling the D-Wave, using the 2000Q lower-noise system, we conclude that the variations in images were not solely due to thermal temperature fluctuations.  However, more experiments are required to prove this claim completely.  Though there were runs when the quantum D-Wave RBM learned only a single digit, we were able to successfully, repeatably create these variations. In addition, we observed when epoch reaches a certain threshold, the samples tend to collapse to a single image. 

Given the samples we collect from the D-Wave after training the RBM using the MNIST dataset encodings and also after training the RBM using the MNIST-Fashion dataset, we show early quantum image generation. 
Thus, paraphrasing Biamonte et. al., "if a small quantum [D-Wave 2000Q] processor can produce statistical patterns that are computationally difficult to be produced by a classical computer, then larger quantum annealers [perhaps the 5000 qubit D-Wave processor using the above hybrid RBM method] might recognize patterns that are significantly more difficult to recognize classically".    

\bibliographystyle{unsrt}
\bibliography{quantum_paper}

\end{document}